\documentclass[11pt]{article}
\usepackage{amsmath}
\usepackage{amsfonts}
\usepackage{amssymb}
\usepackage{graphicx}

\def\be\begin{equation}
 \def\ee{\end{equation}}
\def\bea{\begin{eqnarray}}
\def\eea{\end{eqnarray}}

\begin{document}
\begin{center}
\LARGE {\bf Gravitational Baryogenesis in Anisotropic Universe }
\end{center}
\begin{center}
{\bf  Kh. Saaidi}\footnote{ksaaidi@uok.ac.ir},
{\bf  H. Hossienkhani}\footnote{Hossienhossienkhani@yahoo.com}\\

{\it Department of Physics, Faculty of Science, University of
Kurdistan,  Sanandaj, Iran}

\end{center}
 \vskip 1cm
\begin{center}
{\bf{Abstract}}
\end{center}
The interaction between Ricci scalar curvature and the baryon number current,
dynamically breaks CPT in an expanding universe and leads to
baryon asymmetry. Using this kind of interaction and study the gravitational
baryogenesis in the Bianchi type I universe. We find out the effect of anisotropy
of the universe on the baryon asymmetry for the case which the equation of state
parameter, $\omega$, is dependent to time.\\


\newpage

\section{Introductions}
Theoretical prediction of antimatter  is one of the most impressive
discoveries of quantum field theory which made by Paul Dirac about
80  years ago \cite{1}. Some scientist thought  that "maybe there
exists a completely new universe made of antimatter" because they
believe that there is a symmetry between matter and antimatter.
 Our present point of view on being symmetry between matter and antimatter is very much
different, even opposite.  The absence of $\gamma$ ray emission from
matter- antimatter annihilation \cite{2} and the theory of Big Bang
nucleosynthesis \cite{3}, the measurements of cosmic microwave
background \cite{4}, indicate that there is more matter than
antimatter in the universe. So that,  we  sure that antimatter
exists but believe that there is an asymmetry between matter and
antimatter. The origin of the difference between the number density
of baryons and anti-baryons is still an open problem in particle
physics and cosmology.  Observational results yield that the ratio
of the baryon number to entropy density is approximately $n_b/s \sim
10^{-10}$. The standard mechanism of baryogenesis is based on the
following three principles as it was formulated in 1967 by A. D.
Sakharov \cite{5}:
\begin{enumerate}
\item
 Non-conservation of baryons. It is predicted theoretically by grand unification \cite{6} and even by
the standard electroweak theory \cite{7}.
\item
Breaking of symmetry between particles and antiparticles, i.e.  C
and CP. CP-violation was observed in experiment in 1964 \cite{8}.
Breaking   C-invariance was found earlier immediately after
discovery of parity non-conservation \cite{8a}.
\item
 Deviation from thermal equilibrium. This is fulfilled in nonstationary, expanding universe for
massive particles or due to possible first order phase transitions.
\end{enumerate}
Similarly, in \cite{9}, a mechanism for baryon asymmetry  was
proposed. They  introduced an interaction between Ricci scalar
curvature and any  current that leads to net $B-L$ charge in
equilibrium  ($L$ is lepton number) which dynamically violates CPT
symmetry in expanding Friedmann Robertson Walker (FRW) universe. The
proposed interaction shifts the energy of a baryon relative to an
antibaryon,  giving rise to a non-zero baryon number density in
thermal equilibrium.  The author of \cite{10}
 studied the some mechanism of baryon asymmetry which was proposed in
 \cite{9} for the case which the equation of state parameter of the universe,
  $\omega$, is dependent to  time. Some of another investigations about
   gravitational baryogenesis are done in \cite{10}, \cite{11}, \cite{12},
    \cite{13}, \cite{14}, \cite{15}, \cite{16}.\\
In this paper,  we study the gravitational baryogenesis in the Bianchi
type I universe. We assume the universe is filled  with to components of
perfect fluid and  study this model for different  cases. We will study the effect
of   anisotropy and interaction between two different components of  perfect fluid
  on $\dot{R}$ and consequently on gravitational baryogenesis.

\section{Preliminary}
The gravitational field in our model is given by a Bianchi type I
metric as
\begin{equation}\label{1}
ds^2=-dt^{2}+A^{2}dx^{2}+B^{2}dy^{2}+C^{2}dz^{2},
\end{equation}
where, the metric function, A, B, C, being the function of
time, t, only. \\
We assume the matter is perfect fluid, then the energy momentum
tensor is given by
\begin{equation}\label{2}
T_{\mu\nu}=(\rho+p)u_{\mu}u_{\nu}+pg_{\mu\nu},
\end{equation}
where $u^{\nu}$ is the four vector satisfying
\begin{equation}\label{3}
u^{\nu}u_{\nu}=-1,
\end{equation}
and $\rho$ is the total energy of a perfect fluid and $p$ is the
corresponding pressure. $p$ and $\rho$ are related by an equation
of state as
\begin{equation}\label{4}
p=\omega\rho.
\end{equation}
One can obtain the Einstein field equations form the
BI space time as
\begin{equation}\label{5}
8\pi Gp=-\frac{\ddot{B}}{B}-\frac{\ddot{C}}{C}-\frac{\dot{B}{\dot
C}}{BC},
\end{equation}
\begin{equation}\label{6}
8\pi Gp=-\frac{\ddot{A}}{A}-\frac{\ddot{C}}{C}-\frac{\dot{A}{\dot
C}}{AC},
\end{equation}
\begin{equation}\label{7}
8\pi Gp=-\frac{\ddot{B}}{B}-\frac{\ddot{A}}{A}-\frac{\dot{A}{\dot
B}}{AB},
\end{equation}
\begin{equation}\label{8}
8\pi G\rho=\frac{\dot{A}{\dot B}}{AB}+\frac{\dot{A}{\dot
C}}{AC}+\frac{\dot{B}{\dot C}}{BC},
\end{equation}
where $G$ is the Newtonian gravitational constant and over-dot
means differentiation with respect to $t$.
Using Eqs. (\ref{5}-\ref{8}), we can obtain the Hubble parameter as
\begin{equation}\label{9}
H=\frac{1}{3}\theta=\frac{1}{3}(\frac{\dot{A}}{A}+\frac{\dot{B}}{B}+\frac{\dot{C}}{C})=\frac{\dot{a}}{a},
\end{equation}
\begin{equation}\label{10}
H^{2}=\frac{1}{3}(8\pi G\rho+\sigma^{2}),
\end{equation}
\begin{equation}\label{11}
\dot{H}=-4\pi G(\rho+p)-\sigma^{2},
\end{equation}
\begin{equation}\label{12a}
\dot{\sigma}+\sigma\theta=0,
\end{equation}
where
$a=(ABC)^{\frac{1}{3}}$ is the scale factor, and  $\sigma^2=1/2\sigma_{ij}\sigma^{ij} $  in which $\sigma_{ij}=u_{i,j}+\frac{1}{2}(u_{i;k}u^{k}u_{j}+u_{j;k}u^{k}u_{i})+\frac{1}{3}\theta(g_{ij}+u_{i}u_{j})$ is the shear tensor, which describes the rate of distortion of the matter flow, and $\theta=u^{j}_{;j}$ is the  scalar expansion .
 The equation of state
parameter, $\omega$, can be expressed in terms of the Hubble
parameter  and  shear tensor as
\begin{equation}\label{12}
\omega=-1-\frac{2(\dot{H}+\sigma^{2})}{3H^{2}-\sigma^{2}}.
\end{equation}
We obtain the Ricci scalar as
\begin{equation}\label{13}
R=3H^{2}(1-3\omega)+\sigma^{2}(3\omega-1).
\end{equation}
By derivating of $R$ and use of   Eq. (\ref{11}) and (\ref{12a}), it is shown that
\begin{equation}\label{14}
\dot{R}=\frac{\sqrt{3}}{M_{p}^{3}}(1+\omega)(3\omega-1)\rho\sqrt{\rho+\sigma^{2}M_{p}^{2}}-\frac{3}{M_{p}^{2}}\rho\dot{\omega},
\end{equation}
where, $M_{p}\simeq1.22*10^{19}Gev$ is the Planck mass. If the
space time will be isotropic, $\sigma=0$, Eq. (\ref{14}) reduce to
the result of \cite{10}. Also, if $\dot{\omega}=0$, only the first
term reminds and it is zero at $\omega=1/3$ and at
$\omega=-1$. Therefore by  taking into account $\dot{\omega}$,
we have baryon asymmetry at $\omega=1/3$ and at
$\omega=-1$, because $\dot{R\neq}0$

\section{Perfect Fluid with Interaction.}
In the following we consider our study with universe dominated by
two interacting perfect fluids with equation of states as
\begin{equation}\label{15}
p_{d}=\gamma_{d}\rho_{d},
\end{equation}
\begin{equation}\label{16}
p_{m}=\rho_{m}\gamma_{m}.
\end{equation}
We assume that the  conservation relation of energy for these two components are
\begin{equation}\label{17}
\dot{\rho_{d}}+\theta(\rho_{d}+p_{d})=\Gamma_{1}\rho_{d}+\Gamma_{2}\rho_{m},
\end{equation}
\begin{equation}\label{18}
\dot{\rho_{m}}+\theta(\rho_{m}+p_{m})=-\Gamma_{1}\rho_{d}-\Gamma_{2}\rho_{m}.
\end{equation}
where $\Gamma_{1}\rho_{d}+\Gamma_{2}\rho_{m}$ is the source term
of interaction and $\Gamma_{1}$ and $\Gamma_{2}$ may be time
dependent \cite{17}, \cite{18}, \cite{19}, \cite{20}, \cite{21}.
Although Eq. (\ref{17}), and (\ref{18}) do not satisfy the conservation
equation, but we have
\begin{equation}\label{19}
\dot{\rho}+\theta(1+\omega)\rho=0.
\end{equation}
Here $\rho=\rho_{d}+\rho_{m}$ and $p=p_{d}+p_{m}$ and
\begin{equation}\label{20}
\omega=\frac{\gamma_{d}+\gamma_{m}r}{1+r},
\end{equation}
where $r= \rho_{m}/\rho_{d} $.
Using Eqs. (\ref{15}-\ref{19}), we
obtain
\begin{equation}\label{21}
\dot{\omega}=\frac{\dot{\gamma_{d}}+r\dot{\gamma_{m}}}{1+r}-\frac{(\gamma_{m}-\gamma_{d})(\Gamma_{1}+r\Gamma_{2})}{1+r}-\theta\frac{(\gamma_{m}-\gamma_{d})^{2}r}{(1+r)^{2}}.
\end{equation}
From the third term of Eq. (\ref{21}), it is seen that even for constant equation of  state, $ \omega $ varies with time. This is due to that the universe is supposed to be filled of components with different equation of state parameters.
Substituting  Eq. (\ref{21}) into Eq. (\ref{14}), give
\begin{eqnarray}\label{22}
\dot{R}&=&-\frac{3\rho(\dot{\gamma_{d}}+\dot{\gamma_{m}}r)}{M_{p}^{2}(1+r)}
-\frac{\sqrt{3}}{M_{p}^{3}}\frac{\rho\sqrt{\rho+M_{p}^{2}\sigma^{2}}}{(1+r)^{2}}\cr &\times&
{\Biggr [}(1+\gamma_{m})(1-3\gamma_{m})r^{2}+2(1-\gamma_{m}-\gamma_{d}-3\gamma_{m}\gamma_{d})r+(1-3\gamma_{d})(1+\gamma_{d}){\Biggl ]}\cr
&+&\frac{3\rho}{M_{p}^{2}}\left(\frac{(\gamma_{m}-\gamma_{d})(\Gamma_{1}+\Gamma_{2}r)}{(1+r)}+\frac{\sqrt{3}r(\gamma_{m}-\gamma_{d})^{2}\sqrt{\rho+M_{p}^{2}\sigma^{2}}}{M_{p}(1+r)^{2}}\right).
\end{eqnarray}
We want to check this result for some components.
\subsection {Radiation Dominant}
In this subsection we suppose one of the fluid components correspond to radiation. To do this, we take $\gamma_{m}=1/3$
so that $\dot{\gamma_{m}}=0$ and therefore  Eq. (\ref{22}) reduces to
$$
\dot{R}=\frac{\sqrt{3}}{M_{p}^{3}(1+r)}\rho\sqrt{\rho+M_{p}^{2}\sigma^{2}}(1+\gamma_{d})(3\gamma_{d}-1)-\frac{3\rho\dot{\gamma_{d}}}{M_{p}^{2}(1+r)}$$
\begin{equation}\label{23}+
\frac{\rho}{M_{p}^{2}}\frac{(1-3\gamma_{d})(\Gamma_{1}+\Gamma_{2}r)}{1+r},
\end{equation}
choosing, $\Gamma_{1}=\lambda_{1}\theta$ and $\Gamma_{2}=\lambda_{2}\theta$, $\lambda_{1}$, $\lambda_{2} \in \Re$ \cite{18}, \cite{19}, \cite{20},\cite{21} and one can achieve as
$$
\dot{R}=\frac{\sqrt{3}}{M_{p}^{3}(1+r)}\rho\sqrt{\rho+M_{p}^{2}\sigma^{2}}(1+\gamma_{d})(3\gamma_{d}-1)-\frac{3\rho\dot{\gamma_{d}}}{M_{p}^{2}(1+r)}$$
\begin{equation}\label{24}+
\frac{\rho\theta}{M_{p}^{2}}\frac{(1-3\gamma_{d})(\lambda_{1}+\lambda_{2}r)}{1+r}.
\end{equation}
We assume that the other components which fill the universe, is a massive scalar field of mass $m$, with a time dependent equation of
state parameter
interacting with radiation. The time dependent equation of state parameter of the massive scalar field
as an universe anisotropic universe is define as
\begin{equation}\label{25}
\gamma_{d}=\frac{p_{d}}{\rho_{d}}=\frac{\frac{1}{2} \dot{\phi}^{2}-V(\phi)}{\frac{1}{2}\dot{\phi}^{2}+V(\phi)}.
\end{equation}
Where $V(\phi)=(1/2 )m^2\phi^{2}$.
The interaction between scalar field and radiation is given by Eq. (\ref{17}), and (\ref{18}) with $\gamma_{m}=1/3$. By defined
$z=(1-\gamma_{d})\rho_{d}/2$ and using $\dot{z}=m\rho_{d}\sqrt{1-\gamma_{d}^{2}}$, which was defined in
\cite{22}, we can obtain
\begin{equation}\label{26}
\dot{\gamma_{d}}=-2m\sqrt{1-\gamma_{d}^{2}}+\theta(1-\gamma_{d})[\lambda_{1}+r\lambda_{2}-(1+\gamma_{d})].
\end{equation}
At last by substituting  Eq. (\ref{26}) into Eq. (\ref{24}), we get
\begin{equation}\label{27}
\dot{R}=\frac{6m\sqrt{1-\gamma_{d}^{2}}}{M_{p}^{2}(1+r)}+\frac{2\sqrt{3}}{M_{p}^{3}(1+r)}\rho\sqrt{\rho+M_{p}^{2}\sigma^{2}}(\gamma_{d}-\lambda_{1}-r\lambda_{2}+1).
\end{equation}
For the scalar field dominant, which is equivalent with, $r\longrightarrow 0$, we have
\begin{equation}\label{28}
\dot{R_{\phi}}=\frac{6m\rho\sqrt{1-\gamma_{d}^{2}}}{M_{p}^{2}}+\frac{2\sqrt{3}}{M_{p}^{3}}\rho\sqrt{\rho+M_{p}^{2}\sigma^{2}}(\gamma_{d}-\lambda_{1}+1).
\end{equation}
For the case that $\dot{\phi^{2}} \gg m^{2}\phi^{2}$ ($\dot{\phi^{2}}\ll m^{2}\phi^{2}$) we have $\gamma_{d}=1(-1)$ so that $\dot{\gamma_{d}}=0$ therefore we have
\begin{equation}\label{29}
\dot{R_{\phi}}=\frac{2\sqrt{3}}{M_{p}^{3}}\rho\sqrt{\rho+M_{p}^{2}\sigma^{2}}(2-\lambda_{1}),
\end{equation}
or
\begin{equation}\label{29b}
\dot{R_{\phi}}=-\frac{2\sqrt{3}}{M_{p}^{3}}\rho\sqrt{\rho+M_{p}^{2}\sigma^{2}}\lambda_{1},
\end{equation}
Eq.(\ref{29b}) shows that if there  is no interaction source term with dark matter, i.e. $\Gamma_{1}=0$,  then $\dot{R}_\phi\simeq 0$ and in this case there is no  any gravitational source for asymmetry  in baryon number.
On the other hand for the radiation dominant, i.e., $r\rightarrow \infty$ we obtain
\begin{equation}\label{30}
\dot{R_{R}}=-\frac{2\sqrt{3}}{M_{p}^{3}}\rho_{R}\sqrt{\rho_{R}+M_{p}^{2}\sigma^{2}}\lambda_{2},
\end{equation}
 we see that,  $\dot{R}\neq 0$ if $\lambda_{2}\neq 0$.
It is seen that  isotropic universe, $\sigma=0$, Eq. (\ref{30}) reduce to the result which is obtained in \cite{10}.
We can obtain $\rho_{R}$ as a function of equilibrium temeperatur, $T$. The radiation energy density is related to the
$T$ as \cite{16}, \cite{16a}.
\begin{equation}\label{31}
\rho_{R}=K_{R}T^{4},
\end{equation}
where $K_{R}$ is proportional to the total number of effectively degree of freedom.  So we have
\begin{equation}\label{32}
\dot{R_{R}}=-\frac{2\sqrt{3}}{M_{p}^{3}}K_{R}T^{4} \lambda_{2}\sqrt{K_{R}T+M_{p}^{2}\sigma^{2}}.
\end{equation}

\section{Gravitational Baryogenesis  in Anisotropic Universe}
The author of \cite{9} introduced a mechanism to generate baryon asymmetry. Their mechanism is based on an interaction between
the derivative of the Ricci scalar and the baryon number current, $J^{\mu}$, as
\begin{equation}\label{33}
\frac{\epsilon}{M_{*}^{2}}\int d^{4}x \sqrt{-g}(\partial_{\mu}R)J^{\mu};
\end{equation}
where $M_{*}$ is a cut off characterizing the scale of the energy in the effective theory and $\epsilon=\pm1$.
This interaction  violate CP. The baryon number density in the thermal equilibrium, has been worked out
in detail in \cite{9}.
It lead to
\begin{equation}\label{34}
n_{b}=n_{B}-n_{\bar{B}}=\frac{g_{b}T^{3}}{6\pi^{2}}\left(\frac{\pi^{2}\mu_{B}}{T}+(\frac{\mu_{B}}{T})^3\right),
\end{equation}
where $\mu_{B}$ is a chemical potential and $\mu_{B}=-\mu_{\bar{B}}=-\epsilon\dot{R}/M_{*}^{2}$ and $g_{b}\simeq 1$ is the number of internal degrees of freedom of baryons. According to \cite{16}, the entropy density of the universe is given by $S=(2\pi^{2}/45)g_{s}T^{3}$, where $g_{s}\simeq106$.
The ratio $n_{b}/S$ in the limit $T \gg m_{b}$ and $T\gg \mu_{b}$ is given by
\begin{equation}\label{35}
\frac{n_{b}}{S}=-\epsilon\frac{15g_{b}}{4\pi^{2}g_{s}}\frac{\dot{R}}{M_{*}^{2}T}|_{T_{D}},
\end{equation}
where $T_{D}$ is called the decoupled temperature and in the expanding universe the baryon number violation decouples at the $T_{D}$
temperature. Therefore the baryon asymmetry in terms of temperature can be determined from Eq. (\ref{32}) and (\ref{35}) as
\begin{equation}\label{36}
\frac{n_{b}}{S}\simeq 2.5\frac{\lambda_{2}g_{b}T_{D}^{5}}{\alpha^{2}M_{p}^{5}}+0.04\frac{\lambda_{2}g_{b}\sigma^{2}}{\alpha^{2}(M_{p}T_{D})^{3}},
\end{equation}
where $\alpha=M_{*}/M_{p}$.
\subsection{Baryogenesis with out Interaction}
In this subsection we assume $\gamma_{m}=\gamma_{R}=1/3$.
We assume $\gamma_{d}>1/3$ which is corresponding to non-thermal component and it decrease more rapidly then radiation \cite{9}.
If there is no any interaction between these two component of universe, i.e. $\Gamma_{1}=\Gamma_{2}=0$ then we have
\begin{equation}\label{37}
\dot{\rho_{d}}+\theta(1+\gamma_{d})\rho_{d}=0,
\end{equation}
\begin{equation}\label{38}
\dot{\rho_{R}}+\frac{4}{3}\theta\rho_{R}=0.
\end{equation}
In this case we can arrive at
\begin{equation}\label{39}
\dot{R}=-\frac{\sqrt{3}}{M_{p}^{3}}\frac{\rho\sqrt{\rho+M_{p}^{2}\sigma^{2}}}{1+r}(1+\gamma_{d})(1-3\gamma_{d}),
\end{equation}
and \\
\begin{equation}\label{40}
\dot{r}=r\theta(\gamma_{d}-\frac{1}{3}),
\end{equation}
it is clearly seen that for $\gamma_{d}>1/3$, $\dot{r}>0$. This means that $\rho_{d}$ decreases faster then
$\rho_{R}$. From Eq. (\ref{38}) we can obtain $\rho_{R}\propto (ABC)^{-4/3}\propto a^{-4}$ and then the temperature red shift is as $T\propto a^{-1}\propto (ABC)^{-1/3}$ and also from Eq. (\ref{37}) one can obtain $\rho_{d}\propto T^{3(1+\gamma_{d})}$ \cite{16a}. We suppose at $T=T_{R}$, $\rho_{R}=\rho_{d}$ we have
\begin{equation}\label{41}
\rho_{d}=\epsilon_{R}T_{R}^{4}(\frac{T}{T_{R}})^{3(1+\gamma_{d})},
\end{equation}
\begin{equation}\label{42}
\rho_{R}=\epsilon_{R}T_{R}^{4}(\frac{T}{T_{R}})^{4},
\end{equation}
therefore we obtain
\begin{eqnarray}\label{43}
\frac{n_{b}}{S}&=&\frac{g_{b}(1+\gamma_{d})(1-3\gamma_{d})}{M_{*}^{2}M_{p}}\frac{T_{R}^{2}}{T_{D}}(\frac{T_{D}}{T_{R}})^{1.5(1+\gamma_{d})}\cr &\times&
{\Biggr [}1.28\frac{T_{R}^{4}}{M_{p}^{2}}(\frac{T_{D}}{T_{R}})^{3(1+\gamma_{d})}\sqrt{1+(\frac{T_{D}}{T_{R}})^{1-3\gamma_{d}}}+\frac{0.018\sigma^{2}}{\sqrt{1+(\frac{T_{D}}{T_{R}})^{1-3\gamma_{d}}}}{\Biggl ]}.
\end{eqnarray}\
We assume $T_{D}=\eta T_{R}$. The case which $\eta\gg1$  is equivalence with the state which $r\longrightarrow 0$.  Hence we have
\begin{eqnarray}\label{44}
\frac{n_{b}}{S}&\cong&\frac{(1+\gamma_{d})(3\gamma_{d}-1)g_{b}}{M_{*}^{2}M_{p}}\frac{T_{R}^{2}}{T_{D}}(\frac{T_{D}}{T_{R}})^{1.5(1+\gamma_{d})}\cr &\times&
{\Biggr [}\frac{1.28}{M_{p}^{2}}(\frac{T_{D}^{3(1+\gamma_{d})}}{T_{R}^{3\gamma_{d}-1}})+0.018\sigma^{2}{\Biggl ]}.
\end{eqnarray}\label{45}
For $\eta\ll1$, we can arrive at
\begin{equation}\label{46}
\frac{n_{b}}{S}\simeq(3\gamma_{d}-1)\biggr[\frac{\eta^{4+3\gamma_{d}}}{\alpha^{2}}(\frac{T_{R}}{M_{p}})^{5}+(\frac{\sigma}{\alpha})^{2}\frac{T_{R}}{M_{p}^{3}}\eta^{3\gamma_{d}}\biggl],
\end{equation}
it is clearly seen that if $\gamma_{d}>1/3$, $n_{b}/s>0$.
\section{Conclusion}
The main purpose of the present work has been to explore the consequences of using the anisotropy of metric,
(\ref{1}), as input in Einstein' s equation, assuming that the cosmic fluid is endowed with a perfect fluid. The expression for the energy-momentum tensor $T_{\mu\nu}$ is given in (\ref{2}). The cosmological constant $\Lambda$ has been set equal to zero. We have obtained the following result, for our studies.
\begin{enumerate}
\item
We show  that the universe which dominated by two interacting perfect fluids
has a curvature that varied with time and the effect of anisotropic space time is remarkable.
\item
We have obtained $\dot{R}$ for radiation dominant regime and the effect of anisotropy of space
 time obviously is seen in it.
\item
We assume one of the components which fill the universe, is a
massive scalar field. We have shown that the effect of shear tensor
in $\dot{R}$ is notable and also  for scalar field dominant and for
the case which kinetic term is negligible with respect to potential
term, have obtained $\dot{R}=f(\rho_{R},\sigma)\lambda_1$. This
result shows that in this case, if  $\Gamma_1=0$, there is no any
gravitational source for asymmetry in baryon number.
\item
We have studied the gravitational baryogenesis in anisotropic universe  and have obtained the quantity $n_b/s$ for some  typical example. We have shown that the baryon asymmetry in anisotropic universe is larger than the baryon asymmetry in Friedmann  Robertson Walker (FRW) space time.

\end{enumerate}

\end{document}